\documentclass[twocolumn]{aastex631}

\newcommand{\nubh}{$\nu$\texttt{bhlight}}

\defcitealias{Zaizen2023}{ZN23}
\shorttitle{In-situ flavor transformations}
\shortauthors{Lund et al.}

\usepackage{amsmath}
\usepackage{tikz}
\usepackage{verbatim}

\begin{document}

\title{Angle-dependent in-situ fast flavor transformations in post-neutron star merger disks}

\author[0000-0003-0031-1397]{Kelsey A. Lund}
\affiliation{Department of Physics, University of California, Berkeley, CA 94720, USA}
\affiliation{Institute for Nuclear Theory, University of Washington, Seattle, WA 98195, USA}
\email{klund@berkeley.edu}

\author[0000-0002-3954-2005]{Payel Mukhopadhyay}
\affiliation{Department of Physics, University of California, Berkeley, CA 94720, USA}

\author[0000-0001-6432-7860]{Jonah M. Miller}
\affiliation{Michigan SPARC, Los Alamos National Laboratory, Ann Arbor, MI 48109, USA}
\affiliation{Computational Physics and Methods, Los Alamos National Laboratory, Los Alamos, NM 87545, USA}
\affiliation{Center for Theoretical Astrophysics, Los Alamos National Laboratory, Los Alamos, NM 87545, USA}

\author[0000-0001-6811-6657]{G. C. McLaughlin}
\affiliation{Department of Physics, North Carolina State University, Raleigh, NC 27695, USA}

\begin{abstract}

The remnant black hole-accretion disk system resulting from binary neutron star mergers has proven to be a promising site for synthesizing the heaviest elements via rapid neutron capture (r-process). A critical factor in determining the full r-process pattern in these environments is the neutron richness of the ejecta, which is strongly influenced by neutrino interactions. One key ingredient shaping these interactions is fast neutrino flavor conversions (FFCs), which arise due to angular crossings in neutrino distributions and occur on nanosecond timescales. 
We present the first three-dimensional, in-situ, angle-dependent modeling of FFCs in post-merger disks, implemented within general relativistic magnetohydrodynamics with Monte Carlo neutrino transport. Our results reveal that, by suppressing electron neutrinos, FFCs more efficiently cool the disk and weaken the early thermally driven wind. Less re-leptonization due to electron neutrino absorption makes this cooler wind more neutron-rich, producing a more robust r-process at higher latitudes of the outflow. This study underscores the necessity of incorporating FFCs in realistic simulations.

\end{abstract}

\section{Introduction}
\label{sec:intro}

Detection of the multi-messenger event  GW170817 \citep{LIGO2017a}, has confirmed that binary neutron star (BNS) mergers are indeed sites of rapid neutron capture, or \textit{r-process} nucleosynthesis \citep{Blinnikov1984, Lattimer1976, Lattimer1977,Lippuner2015,Cowperthwaite2017,Villar2017}, but there is as yet insufficient evidence to conclude that BNS mergers alone are responsible for the Galactic inventory of r-process elements.  Thus, a combination of advances in observation and simulation are needed. In dense astrophysical environments, such as those found in the aftermath of neutron star mergers, neutrinos play a pivotal role in mediating matter-radiation interactions \citep{Popham1999,Dimatteo2002} and shaping the composition of outflows \citep[e.g][]{Surman2004, McLaughlin1996, Surman2006, Surman2008, Beloborodov2008, Malkus2012, Caballero2011, Perego2014,Foucart2015,Richers2015, Siegel2017,
Siegel2018,Miller2019,Curtis2022,Curtis2023}.
Among the complex phenomena that emerge in these systems are fast neutrino flavor conversions (FFCs), driven by angular instabilities, known as fast flavor instabilities (FFIs) in neutrino angular distributions.\footnote{For a recent review, see \cite{Johns2025}.} These rapid flavor transformations, occurring on nanosecond timescales, have the potential to dramatically alter the lepton number and electron fraction ($\rm{Y_e}$) within accretion disks, with significant implications for r-process nucleosynthesis. Foundational studies, \citep[e.g.][]{Izaguirre2017, Dasgupta2017, Abbar2018, Morinaga2022} have established that FFIs will necessarily occur wherever the angular distributions of neutrinos and antineutrinos cross each other at a given point in space, a criterion termed the presence of an ELN--XLN crossing. 

\begin{figure*}
    \centering
    \includegraphics[width=0.9\linewidth]{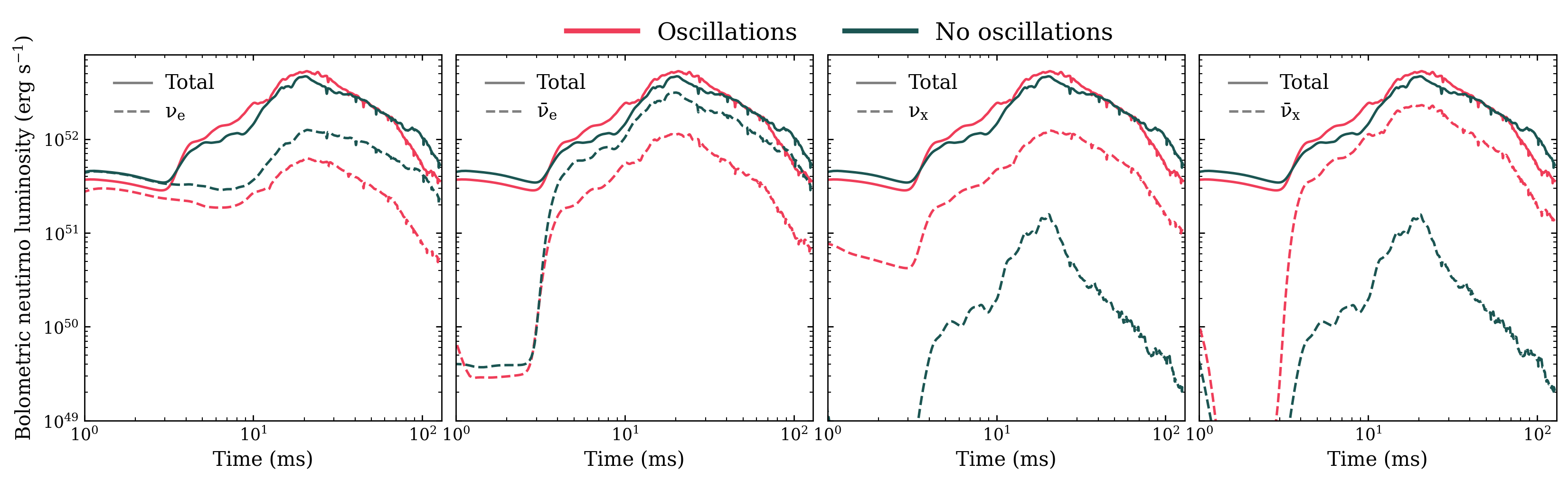}
    \caption{Bolometric luminosity for neutrinos of each species as a function of time for the simulations with (pink) and without (teal) oscillations. From left to right, each panel shows the luminosity from $\nu_e$, $\bar{\nu}_e$, $\nu_{X}$ and $\bar{\nu}_X$as dashed lines, as well as the total neutrino luminosity as solid lines. Oscillations systematically suppress electron flavors and enhance heavy flavors. Furthermore, between $\sim 3$ and $\sim 60$ milliseconds, the total luminosity with FFCs exceeds the luminosity without, suggesting that the disk is more efficiently cooled.}
    \label{fig:nu-lum}
\end{figure*}

The timescale of FFIs is significantly shorter than the typical hydrodynamic timescales in core-collapse supernova (CCSN) or neutron star merger simulations. This vast disparity imposes computational challenges, rendering direct modeling of FFIs in large-scale, multidimensional simulations impractical. To circumvent this complexity, post-processing approaches have been widely employed to evaluate the susceptibility of these systems to FFIs in both CCSN and merger contexts \citep{Wu2017a, DelfanAzari2019, Richers2022a, Mukhopadhyay2024, Kawaguchi2024}. Some studies report the widespread presence of FFIs \citep[e.g.][]{Froustey2024,Wu2017a,Wu2017b,George2020,Richers2022b}, while others reveal intriguing time-dependent structures. For instance, \citet{Mukhopadhyay2024} show that FFIs occur universally at early times but later become localized around the equatorial plane. This result has been broadly corroborated recently in \citet{Kawaguchi2024}. Although such post-processing methods cannot fully capture the true behavior of the neutrino flavor field—since they omit the feedback of neutrino flavor transformations into simulations—they serve as a valuable preliminary step. Such approaches help identify regions where FFIs are most likely to occur, guiding future efforts to incorporate neutrino quantum kinetics into simulations. 

Despite the challenges associated with simulating FFIs dynamically within already computationally intensive simulations, the community has invested tremendous effort in understanding their impact on observables such as nucleosynthesis and the corresponding kilonova.
\citet{Li2021} were the first to dynamically include FFIs in 3D simulations of black hole accretion disks. They demonstrated that FFCs induced by FFIs can enhance neutron-rich outflows, facilitating the synthesis of heavy elements, including lanthanides. Their method relied on classical neutrino transport calculations, identifying regions where instabilities occur, and subsequently modifying the classical neutrino fields using a simplified ad hoc prescription. \cite{Just2022} and \cite{Ehring2023} build on this pioneering work, and follow a similar approach. 

Recently, \cite{Qiu2025} performed a neutron star merger simulation\footnote{Here the end state of the merger is a neutron star, not a black hole.} with a process-agnostic treatment for neutrino flavor transformations, where neutrinos slowly relax towards equipartition, with rates constrained by detailed balance and conservation. They find significant deviation in nucleosynthetic yields and, intriguingly, an impact on the gravitational wave signal.

A major limitation of these works, even the most sophisticated, is a lack of detailed angular information. The presence or absence FFCs is known to depend sensitively on the angular structure of the neutrino field \citep{Johns2021,Mukhopadhyay2024}. Indeed, detailed modeling of neutrino oscillations in a ``zoomed-in'' microphysical context indicate that the final angular structure resulting from flavor transformations is far from simple \citep{Richers2021,Grohs2023,Grohs2024}. As demonstrated by \cite{Miller2019}, this same angular structure also impacts the electron fraction in the outflow and resultant nucleosynthetic yields.

In this \textit{letter}, we present the first fully three-dimensional, angle-dependent, in-situ modeling of FFCs in a post-merger accretion disk. Using GRMHD simulations coupled with Monte Carlo neutrino transport, we track the self-consistent evolution of neutrino flavor states and their impact on disk dynamics. Our results reveal that FFCs strongly influence the neutrino radiation field, shifting the $\rm{Y_e}$ distribution toward lower values and enhancing the neutron-rich conditions necessary for r-process nucleosynthesis.

\section{Methods}
\label{sec:methods}

\begin{figure}
    \centering
    \includegraphics[width=.9\linewidth]{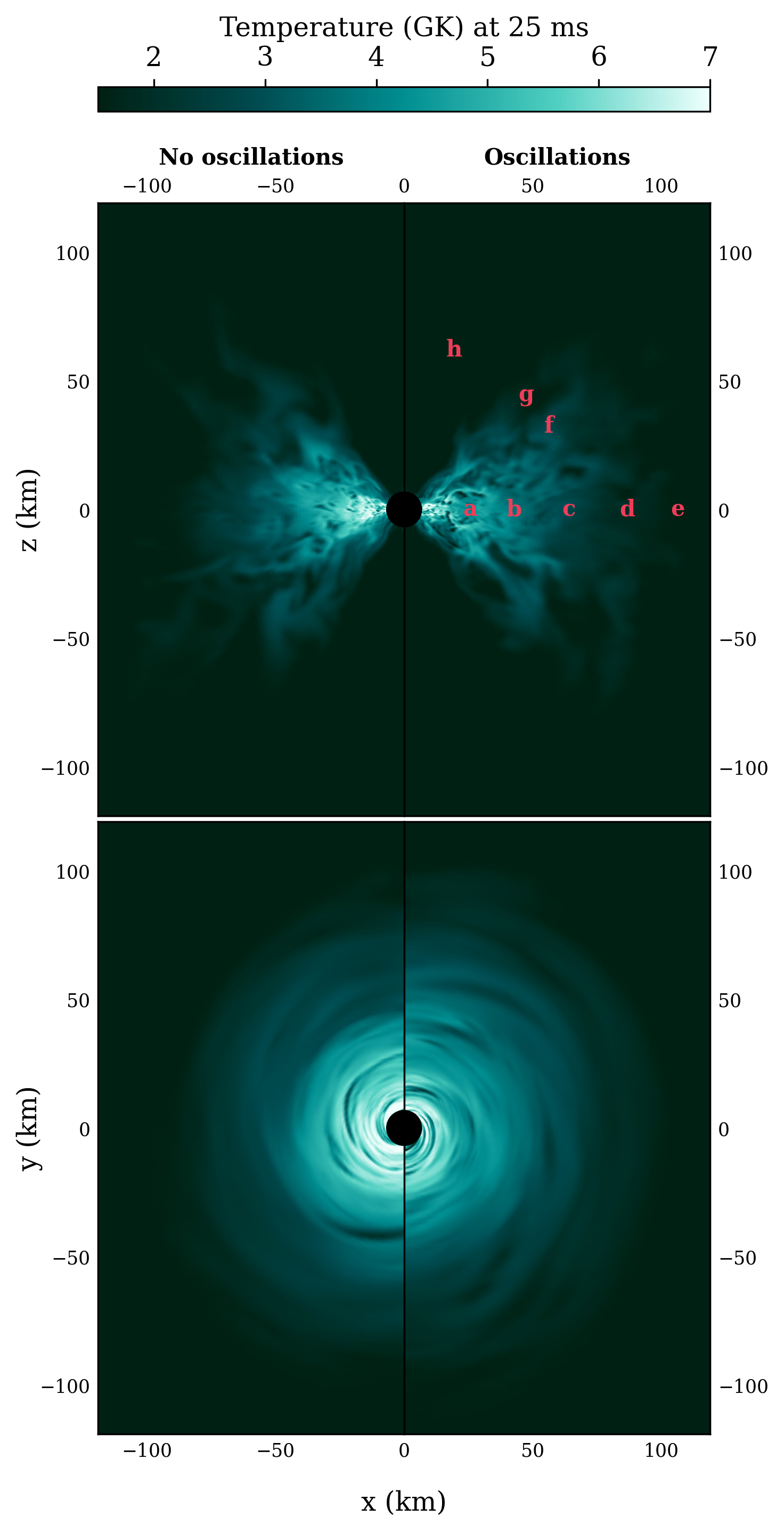}
    \caption{Temperature snapshot at 25\,ms including a side-view (top) and top-down view (bottom). In each case, the simulation with no oscillations is shown on the left, while the simulation with oscillations is shown on the right. FFCs result in a cooler disk, consistent with enhanced neutrino cooling and suppressed thermal wind.}
    \label{fig:T-25ms}
\end{figure}

We solve the equations of general relativistic ideal magnetohydrodynamics (GRMHD) with neutrino radiation transport in a post-merger disk using our code \nubh{}. \nubh{} builds on a long history of methods \citep{Gammie2003,Dolence2009,Ryan2015,nubhlight}. Here we describe details most salient to the current work.

\nubh{} solves GRMHD via finite volumes with constrained transport, and uses Monte Carlo to perform inline neutrino radiation transport. The two are coupled via first-order operator splitting. 
For microphysical data, we use the SFHo equation of state, tabulated in Stellar Collapse format \citep{OConnor2010,OConnor2010_web} and described in \citet{Steiner2013}. For neutrino and opacities, we generate opacity tables via NuLib\footnote{\url{https://github.com/evanoconnor/NuLib}} \citep{NuLib} with interactions from \citet{Bruen1985,Burrows2006}; and \citet{Horowitz2002}.\footnote{We note that the move to NuLib is new to this work. Previous \nubh~models relied on \cite{fornax}.} Scattering is implemented by hand as described in \citet{nubhlight}.

\begin{figure*}
    \centering
    \includegraphics[width=0.9\linewidth]{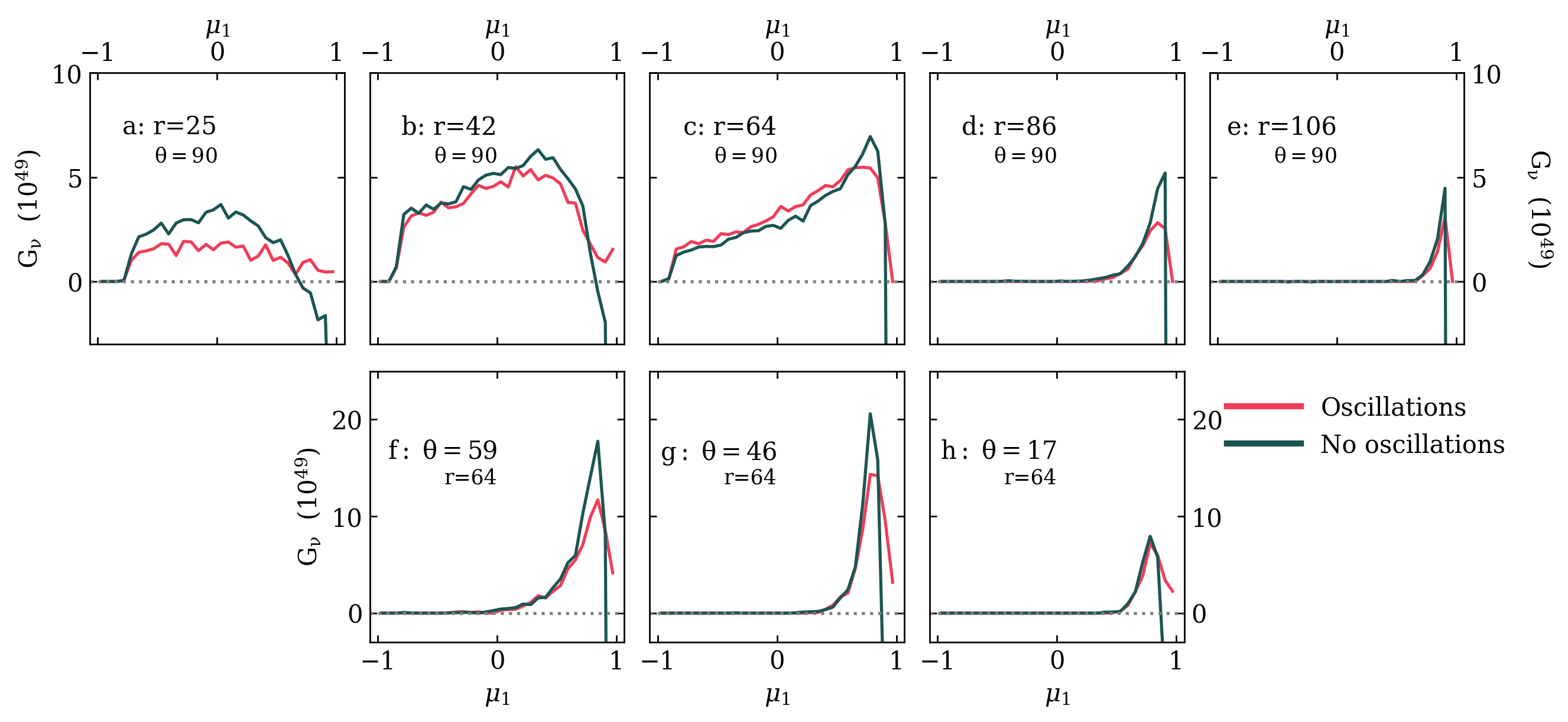}
    \caption{G$_{\nu}$ recorded at 11\,ms, as a function of the angle $\mu_1$, representing the projection of the neutrino wave vector $\vec{k}$ onto the radial unit vector $\hat{r}$, as described in equation~\ref{eq:def:mu1}. Results for two simulations with (pink) and without (teal) oscillations are shown. The radial and polar angle coordinates corresponding to each location are indicated in the top-left corner of each panel (see Fig.~\ref{fig:T-25ms}). FFCs erase ELN--XLN crossings by eliminating the shallower side of the crossing (smaller integral) while conserving the energy-integrated G$_{\nu}$, consistent with expectations.}
    \label{fig:Gnu}
\end{figure*}

We further implement a subgrid heuristic method to capture the FFI. Several approaches have been suggested including \citet{Bhattacharyya2022,Xiong2023b,Abbar2024a,Richers2024}. Our approach is based on the survival probability scheme proposed by \citet[][hereafter \citetalias{Zaizen2023}]{Zaizen2023}, inspired by local quantum-kinetic simulations. A key player in this approach is the crossing indicator function
\begin{equation}
    \label{eq:Gnu}
    G_\nu = (f_{\nu_e} - f_{\bar{\nu}_e}) - \frac{1}{2}(f_{\nu_X} - f_{\bar{\nu}_{X}})
\end{equation}
where here $f_{Q}$ is the phase space density of the neutrino field for species $Q\in\{\nu_e,\bar{\nu}_e, \nu_X,\bar{\nu}_X\}$. We group $\mu$ and $\tau$ neutrinos into the a group of ``heavy lepton'' neutrinos, denoted by the symbol $X$. Unlike in previous \nubh{}~simulations, we do \textit{not} group their antiparticles into the same bin. The antineutrinos for heavy neutrinos are treated separately, which is required to satisfy total lepton number conservation under FFCs.

$G_\nu = 0$ indicates an ELN-XLN crossing. \citetalias{Zaizen2023} argue that fast flavor transformations triggered by an ELN-XLN crossing will serve to eliminate the crossing, while conserving 
\begin{equation}
\label{conservation:law:Gnu}
    \int d\varepsilon d\Omega G_\nu (t, x,y, z,\varepsilon,\Omega)
\end{equation}
where here $\int d\varepsilon d\Omega$ represents an integral over momentum space at fixed position $x,y,z$. To support this, they define the sub-integrals
\begin{eqnarray}
    \label{eq:A}
    A &=& \left| \int_{G_\nu < 0} d\varepsilon d\Omega G_\nu\right|\\
    \label{eq:B}
    B &=& \left|\int_{G_\nu > 0} d\varepsilon d\Omega G_\nu\right|,
\end{eqnarray}
which represent the integrated positive and negative parts of $G_\nu$. 

\citetalias{Zaizen2023} then present the following simple analytic picture for the FFI. When a crossing is present, a flavor transformation occurs. 
The crossing will be eliminated by effectively shrinking the momentum space volume of the regions with the lesser magnitude in their sign until that region vanishes. For example, if $B < A$, then $B$ will be eliminated by shrinking the region of phase space with positive sign.  However, to satisfy conservation law \eqref{conservation:law:Gnu}, the momentum space volume of the larger magnitude region must also shrink.
As a result, during the a FFI event, each neutrino stays in its current flavor with \textit{survival probability}
\begin{equation}
    \label{eq:psurvive}
    p_s = \begin{cases}
    p_{eq} &\text{if in shallower region}\\
    1 - \left(1 - p_{eq}\right) \frac{\min(A,B)}{\max(A,B)} &\text{if in deeper region}
    \end{cases}
\end{equation}
where here $p_{eq}$ is 1/3 for $\nu_e$ and $\bar{\nu}_e$ and 2/3 for $\nu_X$ and $\bar{\nu}_X$.\footnote{This is due to $\nu_X$ bundling two flavors, $\nu_\mu$ and $\nu_\tau$.} We compute these survival probabilities \textit{as a function of space and angle} and, for each neutrino, sample them. For the interested reader, the details of this procedure are described in Appendix \ref{sec:app:nubhlight}. A test problem demonstrating the effectiveness of the method is described in Appendix \ref{sec:onezone:osc}.

\section{Results}
\label{sec:results}

\begin{figure}
    \centering
    \includegraphics[width=\linewidth]{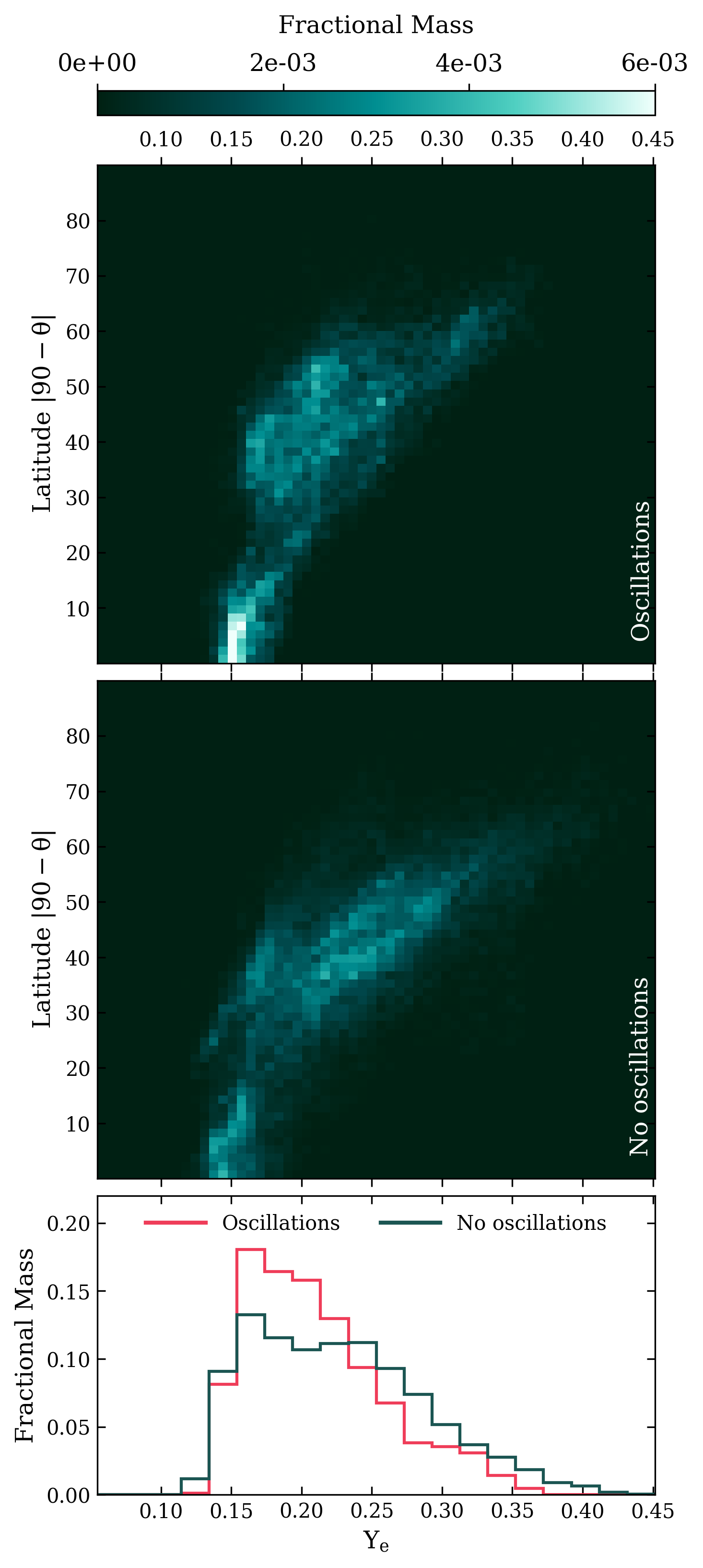}
    \caption{Mass distribution of $\rm{Y_e}$ as the ejecta passes through 10 GK. The top two panels show the distribution of $\rm{Y_e}$ as a function of latitude, with zero representing the plane of the disk. The bottom panel shows the fractional mass distribution for the simulation with (pink) and without (teal) oscillations. The outflow in the simulation with FFCs shifts to lower $\rm{Y_e}$, especially at higher latitudes, indicating enhanced neutron richness in the ejecta.} 
    \label{fig:ye_hists}
\end{figure}

\begin{figure}
    \centering
    \includegraphics[width=\linewidth]{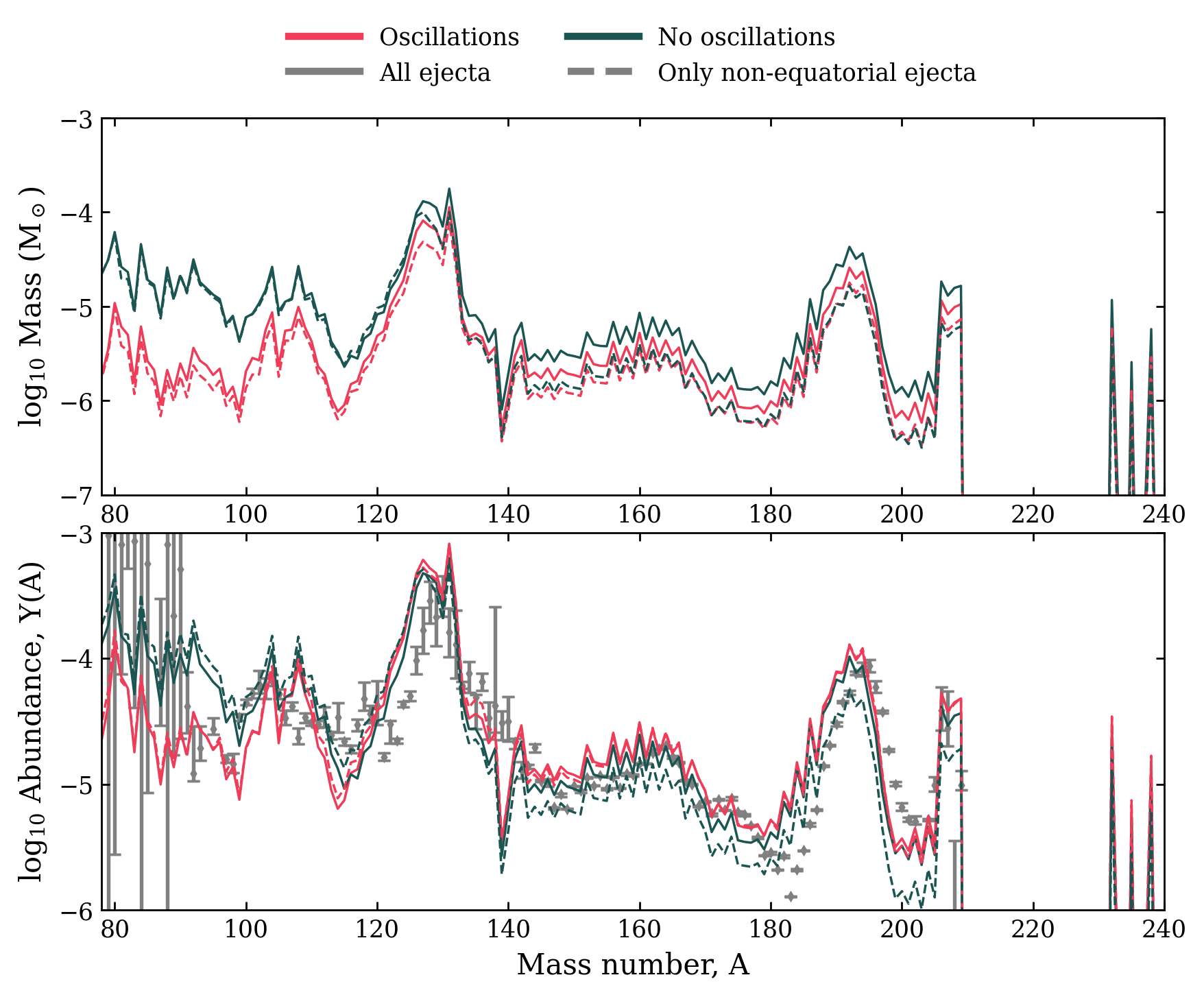}
    \caption{\textit{Top:} Total mass of each element produced, as a function of mass number. \textit{Bottom:} Mass-weighted relative abundances of each element produced, with solar abundances shown as gray diamonds. In each case, we plot both the yields from the outflows with (pink) and without (teal) oscillations. We show the results considering all ejecta (solid lines) as well as non-equatorial ejecta (dashed), as described in Section \ref{sec:results}.
    }
    \label{fig:yields}
\end{figure}

We exercise this capability on a black hole-accretion disk system. We perform two completely new three-dimensional GRMHD simulations of a black hole-accretion disk system, both including four neutrino species: electron neutrinos ($\nu_e$), electron antineutrinos ($\bar{\nu}_e$), and heavy-lepton neutrinos ($\nu_X$, $\bar{\nu}_X$). One simulation, which we denote \textsf{osc}, includes fast flavor conversions (FFCs) while the other, which we denote \textsf{no-osc}, does not. The initial configuration consists of a \cite{Fishbone1976} torus orbiting a \cite{Kerr1963} black hole with dimensionless spin $a = 0.69$ and a disk mass of $0.12M_\odot$. A poloidal magnetic field loop is seeded in the disk, with a gas-to-magnetic pressure ratio of $\beta = 100$ at peak pressure. Magneto-rotational turbulence \citep{Balbus1991} then drives angular-momentum transport and accretion. Resolution details are provided in Appendix \ref{sec:osc:res}.

Figure \ref{fig:nu-lum} shows bolometric luminosities as a function of time for all four flavors for both simulations. The \textsf{osc} simulation shows significantly enhanced $\nu_X$ and $\bar{\nu}_X$ luminosities compared to the \textsf{no-osc}~simulation, and the $\nu_e$ and $\bar\nu_e$ luminosities are commensurately suppressed. Since opacities are highest for electron neutrinos and their antiparticles, this implies that both more electron number and more energy are lost in the \textsf{osc} simulation.

This enhanced loss is reflected in Figure \ref{fig:T-25ms}, which shows the temperature in the disk for both simulations at 25~ms, where the temperature in the \textsf{no-osc} simulation is visibly lower. This reduced temperature provides less heat to drive a thermal wind, which is reflected in the total mass unbound in the ejecta. In the \textsf{osc} simulation, this mass is $1.1 \times 10^{-3} M_\odot$, about half as much as in the \textsf{no-osc} case, which is $2.2 \times 10^{-3} M_\odot$. 

We examine the impact of oscillations on the angular distribution of neutrinos by analyzing the energy-integrated $G_\nu$ at multiple locations in the remnant at 11 ms. In the \textsf{no-osc} simulation, Fig. \ref{fig:Gnu} reveals the presence of crossings at forward angles at various locations, where $G_\nu$ transitions from positive to negative, consistent with the results in \citep{Mukhopadhyay2024}. 
At each crossing, the integrated $G_\nu$ consists of two opposing integrals: a positive contribution on one side and a negative contribution on the other. FFCs eliminate the smaller of these two integrals while adjusting the larger integral to preserve the integrated $G_\nu$. We enforce flavor equilibration on the side of the crossing with the smaller integral; we expect that $G_\nu$ should be zero for in that region the \textsf{osc} case, effectively erasing a crossing that would have existed without oscillations. 
We caution that although crossings present in the teal curves are eliminated in the pink, the two curves represent \textit{different simulations} and cannot be interpreted as ``before'' and ``after.''
Nevertheless, Fig. \ref{fig:Gnu} broadly reflects this effect, with $G_\nu$ in the \textsf{osc} case becoming zero on the side of the smaller integral in the \textsf{no-osc} case. Furthermore, since our method conserves the integrated energy $G_\nu$ at each spatial location, we observe a slight reduction in the positive $G_\nu$ region. 

We analyze the distribution of $\rm{Y_e}$ in the outflow as it cools below 10 GK (see Appendix \ref{sec:app:nuc} for a more detailed description). Figure \ref{fig:ye_hists} shows the histogram of $\rm{Y_e}$ comparing the \textsf{osc} and \textsf{no-osc} simulations. Consistent with the temperature differences seen in Fig. \ref{fig:T-25ms}, we observe that flavor conversions lead to a redistribution of the outflow composition. As electron neutrinos and antineutrinos convert into heavy-lepton flavors, the fraction of lower-$\rm{Y_e}$ material increases relative to the \textsf{no-osc} case. This trend aligns with the decreased neutrino heating and cooler temperatures seen in the left panel of Fig. \ref{fig:T-25ms}, reinforcing the impact of FFCs on the thermodynamic and compositional evolution of the ejecta. The double peak structure visible in the \textsf{no-osc} case in Fig. \ref{fig:ye_hists} was explored in \citet{Lund2024}. The peak at $\rm{Y_e} \sim 0.15$ represents material that was largely equatorially ejected from the disk, while the small second bump around $Y_e \sim 0.20 \, - \, 0.25 $ largely stems from material ejected at an intermediate angle $| 90 -\theta | \gtrsim 30^\circ$. 
These two regions can be seen on the top and middle panels of Fig. \ref{fig:ye_hists}. 
In the \textsf{osc} simulation, we see a shift to a lower $\rm{Y_e}$ and a higher ejection angle.  Although the $\rm{Y_e}$ pattern looks qualitatively similar, these quantitative differences will have consequences for nucleosynthesis.

We determine the nucleosynthesis yields using the r-process nuclear reaction network PRISM \citep{Sprouse2020,Sprouse2021}, following the methods of \citet{Lund2024} and the binning procedure described in Appendix~\ref{sec:app:nuc}. In the top panel of Fig.~\ref{fig:yields}, we show the total mass of each element produced as a function of mass number. The bottom panel displays the mass-weighted relative abundances, overlaid onto the solar r-process abundance pattern for comparison. For both panels, we compare results from the \textsf{osc} and \textsf{no-osc} simulations, showing both the full ejecta (solid lines) and the non-equatorial component ($|90^\circ - \theta| > 30^\circ$, dashed lines). In the \textsf{no-osc} simulation, the full outflow (solid teal line in the bottom panel)—dominated by equatorial ejecta driven by angular momentum transport—produces a robust main r-process. The non-equatorial ejecta (dashed teal line), powered by a thermally driven wind, shows a prominent weak r-process component but a weaker main r-process. When FFCs are included, this non-equatorial ejecta (dashed pink line) becomes more neutron-rich due to suppressed neutrino heating and less re-leptonization, as reflected in the $\rm{Y_e}$ shift shown in Fig.~\ref{fig:ye_hists}. This leads to a suppressed weak r-process component and enhanced production of heavier nuclei. These trends are most visible in the bottom panel, where the shape of the pink dashed curve reflects a diminished weak r-process and a strengthened main r-process as compared with the teal dashed curve.

\section{Outlook}
\label{sec:outlook}
In summary, we present the first in-situ treatment of neutrino flavor transformations that goes beyond the standard assumption of flavor equipartition. We focus on fast flavor conversions (FFCs) and implement a Monte Carlo scheme that samples angular-dependent flavor instability using a physically motivated heuristic proposed by \citetalias{Zaizen2023}. This approach addresses the open question of whether incorporating a more subtle, angle-resolved treatment can meaningfully alter the evolution of post-merger systems. The scheme is particularly appealing: it respects relevant conservation laws while enabling a rich phenomenology in aggregate. Although our implementation is closer to the final state predicted by fine-grained quantum kinetic (QKE) simulations than prior approaches, we still find that FFCs have a substantial impact on r-process nucleosynthesis.

\cite{Lund2024} argue that angle-dependent neutrino transport reveals multiple distinct outflow mechanisms: a low $\rm{Y_e}$ equatorial outflow driven by angular momentum conservation and a higher-$\rm{Y_e}$ thermally driven outflow at higher latitudes. This also appears borne out in \cite{Sprouse2024}. We find the most significant change in the element synthesis pattern due to FFCs is a reduction in weak r-process material in this higher-latitude material. We attribute this effect to a loss of electron neutrinos, leading to a cooler disk and a suppression of the re-leptonization of the wind. On the other hand, we find minimal effect on the main r-process pattern, which is mainly carried by the more equatorial outflow. 

These broad trends belie a rich structure in both the neutrino field and the structure of the disk. Although the big picture agrees with earlier pioneering work, such as \cite{Li2021}, \cite{Just2022}, \cite{Ehring2023}, and \cite{Qiu2025}, our more sophisticated treatment differs quantitatively and in fine details. We leave the exact structure of the neutrino field, and the impact of inline FFCs on it, to future work. For example, there may be structure in the azimuthal direction, due to the relativistic motion of the fluid.  We also leave for future work the inclusion of other types of flavor transformation, e.g. collisional instabilities \citep{Johns2022b,Xiong2023c}, the matter neutrino resonance \citep{Malkus2012,Malkus2014,Zhu2016}, and estimates of the effect of many body correlations \citep{Cervia2019,Balantekin2023,Martin2023,Cirigliano2024}, and ultimately to the inclusion of the quantum kinetic equations, e.g. \cite{Sigl1993,Volpe2013,Vlasenko2014b}. However, this preliminary exploration continues to underscore the importance of angle-dependent neutrino transport and the need to include flavor transformations inline in integrated models. It has implications for many observables such as the expected gamma-ray signal \citep{Korobkin2020,Vassh2024}, the elemental pattern observed in metal-poor stars \citep{Holmbeck2020}, and the late-time slope of the kilonova signal \citep{Zhu2018}.

\begin{acknowledgments}
P.M. is supported in part by the Neutrino Theory Network Program Grant under award number DE-AC02-07CHI11359. 
J.M.M. was supported by the U.S. Department of Energy through the Laboratory Directed Research and Development (LDRD) at Los Alamos under LDRD project 20260598ER. 
This research also used resources provided by the LANL Institutional Computing Program.
LANL is operated by Triad National Security, LLC, for the National Nuclear Security Administration of U.S. Department of Energy (Contract No. 89233218CNA000001).  
This work was supported at NC State by the DOE grant DE-FG02-02ER41216 and DE-SC00268442 (ENAF), by the Office of Defense Nuclear Nonproliferation Research \& Development (DNN R\&D), National Nuclear Security Administration, U.S. Department of Energyand in part under the auspices of the U.S. Department of Energy by Lawrence Livermore National Laboratory with support from LDRD project 24-ERD02.
K.A.L and G.M. acknowledge support from the Network for Neutrinos, Nuclear Astrophysics and Symmetries (N3AS), through the National Science Foundation Physics Frontier Center award No. PHY-2020275.  
This work is approved for unlimited release with LA-UR-25-23002, and the following organizational preprint numbers: INT-PUB-25-005 and N3AS-25-005.

\end{acknowledgments}

\vspace{5mm}

\software{ \nubh{}~\citep{nubhlight},
            NuLib \citep{NuLib},
            HDF5 \citep{HDF5},
            MPI \citep{MPI40},
            OpenMP \citep{Dagum1998_openmp},
            PRISM \citep{Sprouse2020,Sprouse2021},
            Python \citep{VanRossum1995_python},
            NumPy \citep{Harris2020_numpy},
            SciPy \citep{Virtanen2020_SciPy},
            Matplotlib \citep{Hunter2007_matplotlib}
          }

\appendix

\section{Implementation details for our novel neutrino oscillation scheme}
\label{sec:app:nubhlight}

To implement this scheme in a Monte Carlo method like the one in \nubh{}, we require sufficient statistics. In support of this need, we implement the following strategy. We subdivide our radiating region into coarse $NC_1 \times NC_2$ \textit{supercells}. Each supercell spans several cells in $r$ and $\theta$ (in configuration, not momentum space). To further gather statistics, each supercell spans \textit{all of} $\phi$. 

The angular distribution of each neutrino species is then computed, integrated over neutrino energy and the volume of the supercell. In practice, the unit sphere is discretized into angle bins into which neutrinos may be accumulated. This is a two-dimensional grid of angle bins and $G_\nu$ may be computed as a histogram into these angular cells. Since our target application, post-merger disks, is approximately axisymmetric, we assume no crossings will develop in the azimuthal direction. This eliminates one additional degree of freedom in the angular space, allowing us to use only a one-dimensional angular grid, integrated over energy and azimuth. 

As a basis for this one dimensional grid, we project neutrino direction onto the radial unit vector $\hat{r}$ and use
\begin{equation}
  \label{eq:def:mu1}
  \mu_1 = \frac{\hat{r} \cdot \vec{k}}{|\hat{r}||\vec{k}|}
\end{equation}
for wave 3-vector $\vec{k}$. Our angular bins are then evenly spaced in $\mu_1$. As a sanity check, we also project onto a second one-dimensional grid by projecting the neutrino wave vector onto the poloidal unit vector $\hat{\theta}$, producing
\begin{equation}
  \label{eq:def:mu2}
  \mu_2 = \frac{\hat{\theta}\cdot\vec{k}}{|\hat{\theta}||\vec{k}|}
\end{equation}
The angular grids should agree, shifted only by $\pi/2$. And when processing the FFI, we choose whichever grid evidences the deeper crossing. We denote the number of angular grid cells $N_\mu$.

This strategy has the advantage of matching the analysis strategy used by many works in the literature, which also show it is usually sufficient to identify crossings in a disk geometry \citep{Wu2017a,Mukhopadhyay2024}. We emphasize that the classical part of our simulations, the fluid and radiation transport, are fully three dimensional in space for the fluid and six-dimensional in space and angle for the neutrino field. It is only the semiclassical FFI treatment that employs this dimensional reduction.

At low statistics, Monte Carlo noise can introduce unphysical fluctuations in $G_\nu$ which, unless care is taken, may be erroneously interpreted as ELN-XLN crossings. To avoid this situation, we employ two techniques. First, we estimate an ``error'' in $G_\nu$ by estimating the variance of the distribution function. Conceptually, this variance is the shot noise in a given bin in phase space. In practice the absolute error takes the form
\begin{equation}
\label{eq:stddev}
    \sigma^2 
    =N_s \left\langle w\right \rangle^2 + \frac{N_s}{N_s-1}\left[\left(\sum_{p\in \text{bin}} w_{p}^2\right) - N_s \left\langle w\right\rangle^2\right]
\end{equation}
where here $w_p$ is the weight per packet $p$, $$\left\langle w\right\rangle = \sum_{p\in\text{bin}} w_p / N_s$$ is the mean weight. Sums are over all packets in an angular bin in a supercell and $N_s$ is the total number of packets in a given bin.

Additionally we estimate the time scale over which the FFI takes place as \citep{Froustey2024}
\begin{equation}
  \label{eq:tau:osc}
  \tau_{FFI} = \frac{\hbar}{\sqrt{2} G_F n}
\end{equation}
where here $\hbar$ is Planck's constant, $G_F$ is Fermi's constant and $n$ is the number density of neutrinos at a point in space. We compute the minimum $\tau_{FFI}$ over the simulation domain and only activate our oscillation machinery when it is less than a single radhydro time step, which is set by the light crossing time of the smallest cell in our simulation. For a post-merger remnant, this is about 100ns.

\section{Toy one-zone oscillation problem}
\label{sec:onezone:osc}

\begin{figure*}[t]
    \centering
    \includegraphics[width=0.9\linewidth]{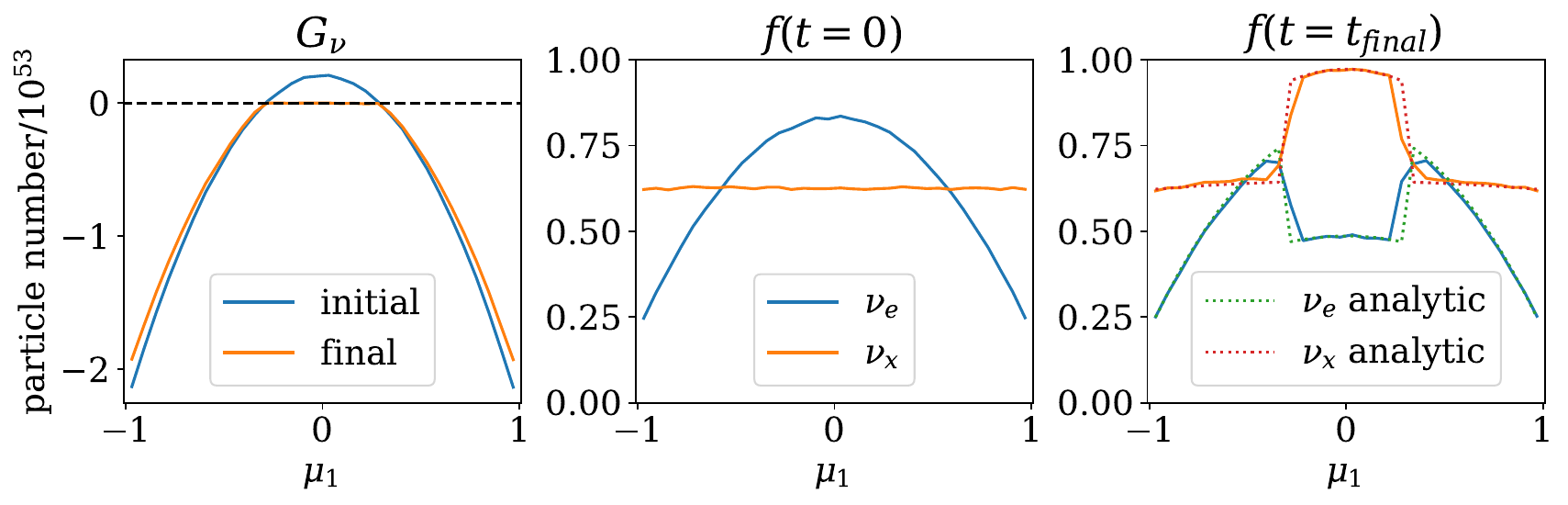}
    \caption{One-zone oscillation problem demonstrating efficacy of our scheme. Central panel shows angular distribution of electron and heavy neutrinos. Right panel shows final configuration, overlaid with the analytic solution. Left panel shows initial and final values of $G_\nu$. We initialize a neutrino field with a crossing and our approach removes the crossing while conserving integrated $G_\nu$.}
    \label{fig:toy}
\end{figure*}

To demonstrate the effectiveness of our strategy we present the following test problem. We initialize a constant number $N_\nu$ of neutrinos represented by $N_{MC}$ Monte Carlo packets each with equal weight, with an artificial spatially homogeneous neutrino distribution in Minkowski space. All neutrinos are initialized with identical energy $E$ and a wave-vector lying in the $x-y$ plane with their angle off of the $\hat{x}$ unit vector specified by projection angle $\mu = \vec{k} \cdot \hat{x}$. $\mu$ is randomly sampled from a probability distribution $f_0(\mu)$ that depends on neutrino species:
\begin{equation} 
\label{eq:toy:f:init} 
f_0(\mu) = \begin{cases}
  \frac{2}{3}\left[1 - \frac{3}{4} \mu^2\right] &\text{for }\nu_e\\
  \frac{1}{4} - {3}{4} \mu^2 &\text{for }\bar{\nu}_e\\
  \frac{1}{2}&\text{for }\nu_x\text{ and }\bar{\nu}_x
\end{cases}.  
\end{equation} 
The fractions of each flavor are: $1/5$ for $\nu_e$, $2/5$ for $\bar{\nu}_e$, $1/5$ for $\bar{\nu}_X$, and $1/5$ for $\bar{\nu}_X$. 

According to \citetalias{Zaizen2023}, the phase space density after the FFI saturates should be given by
\begin{equation}
    \label{eq:toy:f:final:nue}
    f_{\nu_e} = \begin{cases}
      \left(1 - \frac{2}{3}\frac{B}{A}\right) f_{0,\nu_e} + \frac{1}{3}\frac{B}{A}f_{0,\nu_x}&\text{in region }A\\
      \frac{1}{3}\left(f_{0,\nu_e} + f_{0,\nu_x}\right)&\text{in region }B
    \end{cases}
\end{equation}
for $\nu_e$ and
\begin{equation}
    \label{eq:toy:f:final:nux}
    f_{\nu_x} = \begin{cases}
      \frac{2}{3}\frac{B}{A}f_{0,\nu_e} + \left(1 - \frac{1}{3}\frac{B}{A}\right)f_{0,\nu_x}&\text{in region }A\\
      \frac{2}{3}\left(f_{0,\nu_e} + f_{0,\nu_x}\right)&\text{in region }B
    \end{cases}
\end{equation}
for $\nu_x$. Here $f_{0,q}$ represents the initial phase space density for species $q$.

For the plots described below, we choose $N_{MC}=10^7$ and $N_{\nu}=10^{55}$. However, smaller $N_{MC}$ still produces reasonable answers. We choose a number of angular grid cells $N_\mu=32$. The spatial resolution for this problem is irrelevant, as it is homogeneous. We use $N_1\times N_2\times N_3 = 12^3$ spatial grid cells and $NC_1\times NC_2 = 12^2$ supercells.

The initial distribution functions as a function of $\cos(\theta)$\footnote{All other independent variables are integrated out.} for $\nu_e$ and $\nu_x$ are shown in the middle panel of Figure \ref{fig:toy}, and the final distributions are shown in the right panel. The analytic solution given by equations \eqref{eq:toy:f:final:nue} and \eqref{eq:toy:f:final:nux} are overlaid as dotted lines. Due to the finite bin width of the angular bins, the numerical solution does not quite agree with the analytic at exactly the crossing points between the initial $A$ and $B$ regions. However, other than this edge effect, agreement between analytic and numerical solutions is excellent.
The initial and final values of $G_\nu$ are shown in the left panel. Our procedure eliminates the crossing, subtracting $G_\nu$ from the shallower region while conserving $G_\nu$ as a whole. Numerically, The integral over $\mu$ of $G_\nu$ is satisfied up to shot noise, $1/\sqrt{N_{MC}}$. Here, $G_\nu$ is conserved up to one part in $10^3$.

\begin{figure*}[t]
    \centering
    \includegraphics[width=0.9\linewidth]{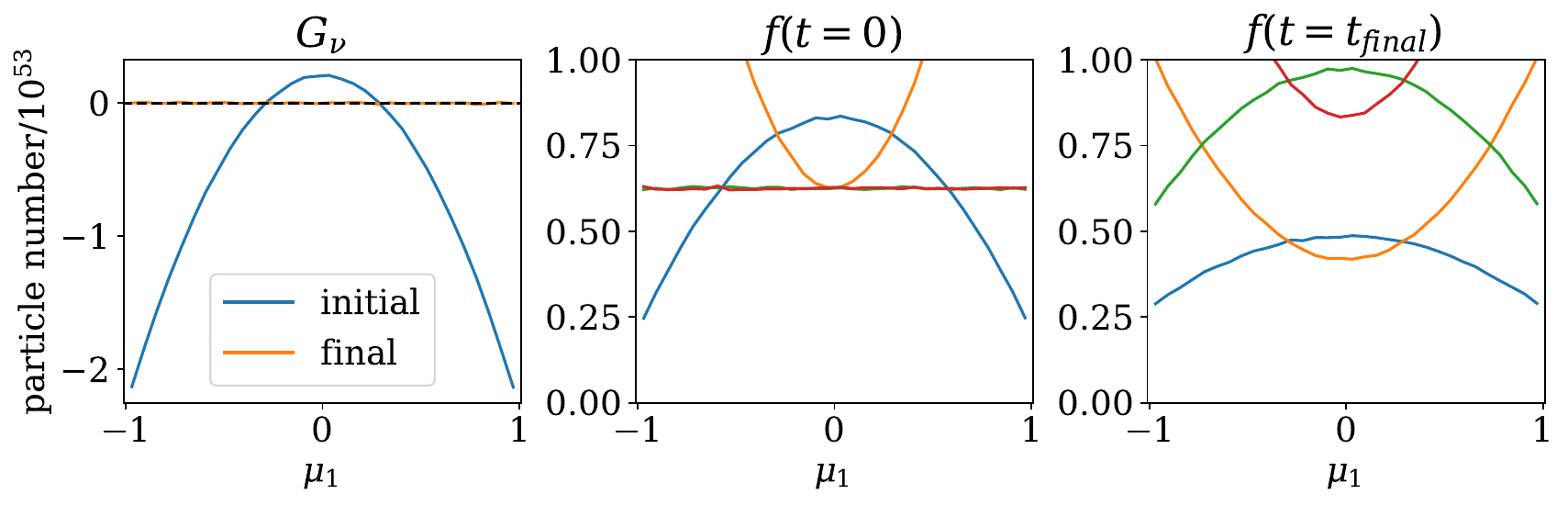}
    \caption{Same as Figure \ref{fig:toy} but where we \textbf{incorrectly} enforce equipartition across all angles of the superzone. In the middle and right panels, we now show all four species. In this procedure, the crossing is fallaciously eliminated by setting $G_\nu=0$ everywhere. The resulting distributions for the four species are significantly distorted.}
    \label{fig:toy:bad}
\end{figure*}

To demonstrate the importance of tracking angle in eliminating a crossing, we also run the same problem, but when a crossing is detected, we simply force all species into equipartition with each other (while still conserving total lepton number). This erroneous procedure is essentially the only option available to moment methods, such as in \citet{Li2021} and \cite{Qiu2025}, where the full angular distribution is not available. Moment methods may miss or incorrectly identify ELN-XLN crossings due to this same limited access to angular information.
 We show the results of applying this incorrect equiparition method in Figure \ref{fig:toy:bad}. In this case, rather than showing the analytic distribution, we show the antiparticles. Here, the crossing is eliminated by setting $G_\nu=0$ everywhere in the supercell with a crossing. While total lepton number is still conserved, the distributions of all species are significantly distorted. In particular, the heavy neutrinos and their antiparticles gain significant number, especially in the initially deeper $A$ region, at the cost of the electron neutrinos and their antiparticles.

\section{Resolution requirements}
\label{sec:osc:res}

In a production \nubh{}~simulation, we typically run with a spatial grid of $N_r\times N_\theta \times N_\phi = 192\times 128 \times 66$ zones, or $\sim 10^6$ zones total, with about 10 particles per cell or a total of $10^7$ Monte Carlo packets total. With high-order reconstructions and the FMKS coordinate system described in \citet{McKinney2004}, this is sufficient to resolve the fastest growing mode of the magneto-rotational instability for our initial conditions and satisfy the stability requirements for Monte Carlo, as well as accurately capture the neutrino field. For more details on these solution requirements and how an adequate resolution is determined, see the appendix of \citet{Miller2019} and the relevant sections of \citet{Porth2019}.
Experiments with the toy problem described in Section \ref{sec:onezone:osc} indicate we require about $10^5$ particles per supercell and at least $N_\mu = 32$ angular bins. To gather enough particles per supercell, without changing the particle number we run at, we use $NC_1\times NC_2 = 32^2$ supercells. We also add the supercells only to the \textit{radiatively active} region of the simulation, where neutrinos are not free streaming. This excludes a large fraction of the domain.

\section{Detailed nucleosynthesis procedure}
\label{sec:app:nuc}
The thermodynamic evolution of the fluid is recorded via approximately $2\times 10^6$ Lagrangian tracer particles, as described in Sec. 3.6 of \citet{nubhlight}. We perform an analysis as described in Sec. 2 of \citet{Lund2024} whereby we identify tracers that passed through a radius of at least 250 gravitational radii ($\sim 10^4$ km) by the end of the simulation and have a Bernoulli parameter, $B_e>0$, of which there are about $10^{5.5}$. We then record the properties of each tracer the last time it drops below T=10~GK, which we refer to as $t_{10}$ in \citet{Lund2024}, signifying the transition out of nuclear statistical equilibrium.

We reduce the volume of tracers required to analyze the nucleosynthetic yields via novel procedure, similar to that described in \cite{Radice2016}. Given that the electron fraction plays a large role in determining the nucleosynthetic outcome of a given trajectory, we sample $\rm{Y_e}$ in the disk outflow to obtain a reduced set of tracers. For each of 15 $\rm{Y_e}$ bins, we select two tracers: one sampled from fast-moving material and one from slow-moving material, with a radial velocity of 0.08c as the cut-off between the two categories. This results in $15 \times 2 $ tracers representing the ejecta. The mass of each tracer is then rescaled for the purpose of post-processing such that it equals the mass contained in the bin it represents. 
To obtain a similar sample for the non-equatorial material, we perform the same analysis excluding tracers whose position at $t_{10}$ lie above 30\textdegree above the mid-plane. 

For each tracer in each sample, we use the Portable Routines for Integrated nucleoSynthesis Modeling \citep[PRISM][]{Sprouse2021} to compute abundances at 1 Gyr post-merger. We use the same nuclear data as in \cite{Lund2024}. Namely, we use the JINA Reaclib library \citep{Cyburt2010} for charged-particle and light-nuclei interactions, theoretical beta decay rates from \citet{Moller2019}, and beta-delayed neutron emission and fission probabilities using \citet{Mumpower2016}.
Neutron capture and neutron-induced fission rates use CoH \citep{Kawano2016}. We use the barrier height-dependent prescription from \citet{Karpov2012} and \citet{Zagrebaev2011} to calculate spontaneous fission rates using the FRLDM barrier height description \citep{Moller2015}. Our theoretical alpha decay rates come from the Viola-Seaborg relation. Finally, where experimental or evaluated data exists, we overwrite theoretical values with data from the 2020 version of NuBase \citep{Kondev2021} and the Atomic Mass Evaluation \citep{Wang2021}.

\bibliography{manuscript}{}
\bibliographystyle{aasjournal}

\end{document}